# Dissipative Topological Defects in Coupled Laser Networks

Vishwa Pal, Chene Tradonsky, Ronen Chriki, Asher A. Friesem and Nir Davidson

Department of Physics of Complex Systems, Weizmann Institute of Science, Rehovot 76100, Israel.

vishwa.pal@weizmann.ac.il

**ABSTRACT**

Topologically protected defects have been observed and studied in a wide range of fields, such as cosmology, spin systems, cold atoms and optics as they are quenched across a phase transition into an ordered state. Revealing their origin and control is becoming increasingly important field of research, as they limit the coherence of the system and its ability to approach a fully ordered state. Here, we present dissipative topological defects in a 1-D ring network of phase-locked lasers, and show how their formation is related to the Kibble-Zurek mechanism and is governed in a universal manner by two competing time scales of the lasers, namely the phase locking time and synchronization time of their amplitude fluctuations. The ratio between these two time scales depends on the system parameters such as gain and coupling strength, and thus offers the possibility to control the probability of topological defects in the system. Enabling the system to dissipate to the fully ordered, defect-free state can be exploited for solving hard combinatorial optimization problems in various fields. As opposed to unitary systems where quenching is obtained via external cooling mostly through the edges, our dissipative system is kept strictly uniform even for fast quenches.

**INTRODUCTION**

Topological defects have been extensively studied in cosmology, spin systems and optics [1]. Their origin and scaling behavior was first explained by Kibble-Zurek (KZ) mechanism [2, 3], which was experimentally studied in various systems, such as atomic gases [4-9], nonlinear optics [10], and condensed matter systems [11-13]. Due to complexity and experimental limitations, these systems do not fulfil the exact KZ mechanism, and only deal with limited aspects [5, 11, 14, 15]. The formation of defects in the Zurek's approach relies on the notion of competing time scales in the system, and the density of defects follows a power-law behavior with respect to the rate at which the phase transition is crossed [13]. The topologically protected defects may prevent the system from reaching a globally stable state with spontaneous symmetry breaking and long range ordering [8]. In most of these works the phase transition into an ordered state was crossed by cooling the system from the outside, whereby the external cooling rate determined the density of defects. For the most interesting regime of fast cooling, it is hard to maintain uniformity over the entire system [4].

Here we present and characterize a new mechanism to form topological defects without external cooling in a dissipative system of coupled laser network [16-18]. In coupled lasers and polaritons networks, losses depend on the relative phase between all oscillators [18-21]. This provides dissipative coupling [20] that can drive the system to a stable steady state phased-locked solution with minimal loss that can be directly mapped to the ground state of the classical XY spin Hamiltonian [18, 19, 21]. However, when the dissipative dynamics is highly over-damped [22] and occurs on a complex landscape, the system fails to reach the globally stable solution and gets stuck in local minima. For a 1-D system on a closed ring, such local minima are topological defects [8], characterized by a non-zero phase circulated over the ring that must be an integer multiple of $2\pi$ (winding number or topological charge) [23]. The degeneracy of the co-existing degenerate solutions can further leads to frustration [18].

We study such topological defects on a 1-D laser ring network with nearest neighbor dissipative coupling. Such laser networks can be approximated as networks of Kuramoto phase oscillators [25] under the assumption of constant intensities, and it is known that such networks cannot unwind topological defects [24]. As we will show both experimentally and numerically, the parameters of our system can be tuned to control the probability of finding topological defects in the network at steady state. When the lasers are pumped hard the system approaches the limit of Kuramoto phase oscillators where topological defects are stuck, while when the lasers are only weakly pumped the system can unwind topological defects. The mechanism of unwinding topological defects is identified as fluctuations of the laser field amplitudes, which are coupled to the laser phase dynamics, and anneal the system to the globally stable uniform phase (defect-free) state. According to this picture, the amplitude fluctuations act as an effective temperature of the system. At high pump strengths, amplitude fluctuations die out very fast and the topological defects don't have enough time to unwind – just as topological defects would be stuck in a thermodynamic system if the system would be cooled too quickly. Recently, a nice analogy between the temperature and pump strength was demonstrated, where the threshold pump strength was shown to link with the absolute zero temperature [25].

The formation of dissipative topological defects is also connected to competition between time scales, as in the KZ mechanism. In laser networks, the competing time scales are the phase locking time [26, 27] (equivalent to the *internal* information exchange time in KZ) and synchronization time of the lasers amplitude fluctuations [28] (equivalent to the *external* cooling rate in KZ). These two time scales are controlled by many system parameters such as the pump and coupling strength but the defect formation can be expressed by single universal dependence on the ratio between them.

Our approach to find the globally stable, minimum loss solution (and avoid getting stuck in local minima) can be exploited for solving combinatorial optimization problems in various fields such as social network, artificial intelligence, wireless communications, biology and medicine [29]. Computational solutions for such optimization problems are often NP-hard [30], so intensive research has been focused on using alternative physical systems. Recently, few non-deterministic polynomial time hard problems using degenerate optical parametric oscillators (DOPOs) as an Ising machine were demonstrated [25, 31, 32]. A large-scale geometric frustration in the Kagome lattice was also demonstrated using thousands of coupled lasers [18].

**EXPERIMENTAL ARRANGEMENT AND NUMERICAL MODELS**

The experimental arrangement for investigating and controlling topological defects, using 1-D coupled lasers network, is schematically presented in Fig. 1, along with representative results. The network of lasers is generated in a degenerate cavity which consists of two cavity mirrors, a 4f telescope, a mask containing circular holes 10–30 in a ring geometry, and a Nd:YAG gain medium pumped by a 100 μs pulsed xenon flash lamp. The intra-cavity 4f telescope ensures that any field distribution at the mask plane is imaged onto itself after every round-trip. Accordingly, each hole on the mask act as an independent individual laser [23]. Coupling between adjacent lasers is introduced by moving the mirror $M_1$ away from the mask a distance d, whereby light diffracted from each laser couples into neighboring lasers resulting in 1-D ring network of lasers. For efficient phase locking, this distance d was set to quarter of Talbot length [33], which provides negative coupling between the adjacent lasers, and results in a solution of out-of-phase (0, π, 0, π, …) locked lasers [34]. The output from the lasers was then focused on to a nonlinear KTP crystal, so both the frequencies as well as the phases are doubled, and then imaged on CCD camera. In such second harmonic generation (SHG) scheme, the out-of-phase solution is converted into an in-phase solution (0, 0, 0, 0, …), which is equivalent to phase locking of lasers with positive coupling [34]. The conversion from out-of-phase to in-phase enables quantitative analysis of globally stable and locally stable solutions.

Representative experimental results when phase locking 10 lasers on a ring network are shown in Fig. 1(b). The dark center of the first harmonic far-field intensity distribution indicates out-of-phase locking, whereas the bright central peak of the second harmonic far-field intensity distribution denotes in-phase locking of the lasers, in excellent agreement with calculated results [35]. Henceforth we only consider second harmonic results that correspond to positive coupling between adjacent lasers.

While a continuous system can have continuum of stable solutions, a discrete network has a finite number [36]. For example, as shown in Fig. 1(c), a ring of 10 coupled lasers has 9 steady-state solutions, where the phase of the laser m is $\varphi_m = e^{iqm2\pi/10}$ and q = -4, -3, -2, -1, 0, +1, +2, +3, +4 is the winding number (topological charge) of each solution. The q=0 in-phase solution is globally stable, whereas the eight q≠0 helical phase solutions are only locally stable. The dissipative coupling between the lasers causes the system to converge to steady-state solutions with minimal loss [20]. The solutions with lower $|q|$ have lower losses, so they are more probable in steady-state. The difference in relative losses between the globally stable and locally stable solutions also reduces with the system size (number of lasers) [24].

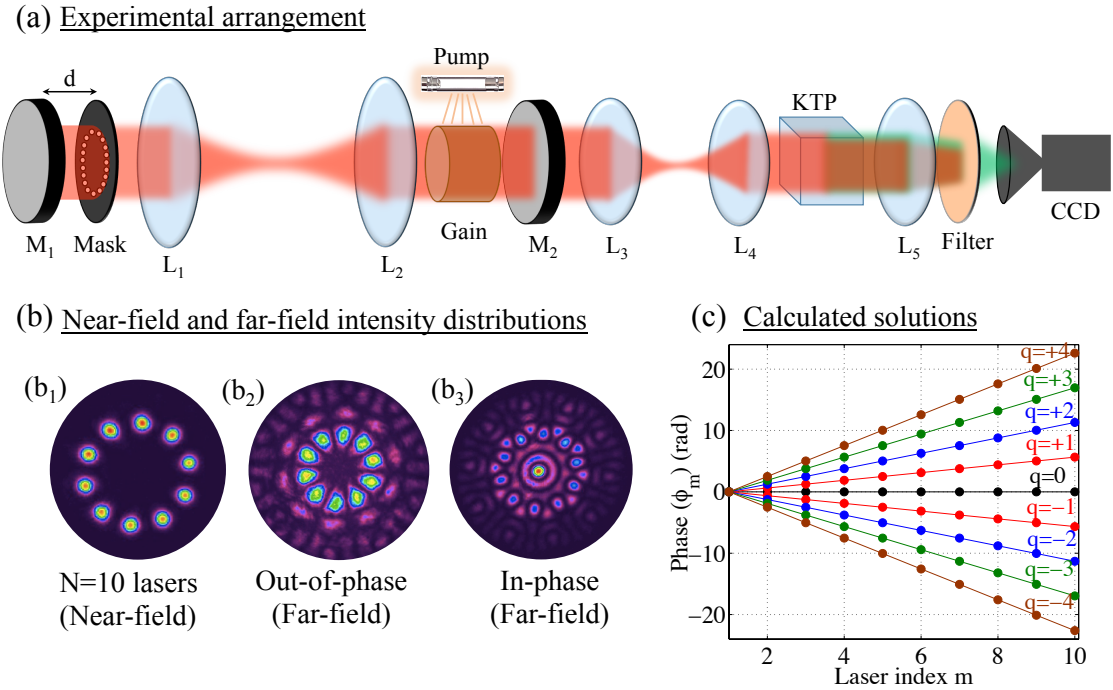

FIG. 1 Dissipative topological defects in a 1-D ring network of coupled lasers. (a) Experimental arrangement; $M_1$ (high reflectivity) and $M_2$ (93% reflectivity) are cavity mirrors, $L_1$ and $L_2$ are plano-convex lenses in a 4f telescope configuration, $L_3$ and $L_4$ are plano-convex lenses in a demagnifying telescope configuration, KTP is a nonlinear crystal for second harmonic generation, $L_5$ is an imaging lens. The filter transmits only frequency-doubled wavelength. (b) Representative experimental results for a ring network of 10 lasers: ($b_1$) near-field intensity distribution, ($b_2$) first harmonic far-field intensity distribution, and ($b_3$) second harmonic far-field intensity distribution. (c) Calculated steady-state solutions for different winding numbers q = -4, -3, -2, -1, 0, +1, +2, +3, +4.

For ring laser networks to converge to a stable solution, all lasers must be phase locked and fulfill periodic boundary conditions. Yet, according to Mermin-Wagner theorem long-range order cannot exist in one-dimensional systems in the thermodynamic limit [37, [38]]. Consequently, we limited our investigation to networks with ≤ 30 lasers, and ensured that all lasers are phase locked using a direct interference analysis [23, 35].

To support and verify our experimental results, we numerically solved the rate equations that characterize a set of N single transverse and longitudinal modes class-B lasers, which are coupled linearly to each other through their optical fields [39], as

$$\frac{dA_m}{dt} = \frac{1}{\tau_p}(G_m - \alpha_m)A_m + \sum_{n=(m)_{NN}}^{N} \frac{\kappa_{mn}}{\tau_p} A_n \cos(\varphi_n - \varphi_m), \qquad (1)$$

$$\frac{d\varphi_m}{dt} = \Omega_m + \sum_{n=(m)_{NN}}^{N} \frac{\kappa_{mn}}{\tau_p} \frac{A_n}{A_m} \sin(\varphi_n - \varphi_m), \qquad (2)$$

$$\frac{dG_m}{dt} = \frac{1}{\tau_c}(P_m - G_m(|A_m|^2 + 1)). \qquad (3)$$

$A_m$, $\phi_m$, $G_m$, $\alpha_m$, $\Omega_m$, and $P_m$ are the amplitude, phase, gain, loss, frequency detuning and pump strength of laser m, $\tau_p$ denotes the cavity round-trip time, $\tau_c$ is the carrier lifetime, $\kappa_{mn}$ the coupling coefficient between lasers m and n is the normalized overlap of the field of laser m and the field of laser *n* after one round trip propagation, and $n = (m)_{NN}$ is the sum over laser m's NN (nearest neighbors) [40]. The coupling between the next and higher order neighbor lasers were negligible in our system. The ring geometry was accounted by the periodic conditions $E_{m+N} = E_m$, where $E_m = A_m e^{i\varphi_m(t)}$. The coupled equations (Eqs. (1)-(3)) were numerically solved using fourth order Runge-Kutta (RK) method, using our system parameters of $\tau_p$ = 5.4 ns, $\tau_c$ = 230 μs, $\alpha_m$ = 0.1. The distribution of detuning was $\Omega_m \ll \kappa_{mn}/\tau_p$, as provided by our degenerate cavity [18], ensuring that the coupling strength was well above the critical value [28, 26]. In the simulation, the coupling value was taken as $\kappa_{mn}$ = 0.01.

If the amplitudes $A_m$ of all lasers are equal, the dynamics of the lasers phases decouple from the field amplitudes, and Eq. (2) results in the well-known coupled Kuramoto phase oscillators [22, 18], as

$$\frac{d\varphi_m}{dt} = \Omega_m + \sum_{n=(m)_{NN}}^{N} \frac{\kappa_{mn}}{\tau_p} \sin(\varphi_n - \varphi_m). \qquad (4)$$

The Kuramoto model was shown to successfully model a large class of problems such as populations of coupled oscillators [22], complex networks [41] and coupled laser arrays [42, 18].

**EXPERIMENTAL AND NUMERICAL RESULTS**

We quantified topological defects by analyzing the steady-state far-field intensity distributions measured after SHG, as described above. The results for ring networks of 10 and 20 lasers are presented in Fig.2. For 10 lasers (Figs. 2(a)-2(c)), the far-field intensity distribution has clearly observable bright central peak and distinct dark and bright rings, indicating a nearly pure q=0 in-phase solution. For 20 lasers (Figs. 2(d)-2(f)), the far-field intensity distribution has largely smeared dark and bright rings, manifesting the presence of topological defects [23]. Due to the multiplicity of longitudinal modes in each laser, each experimental realization corresponds to an ensemble average of many independent experiments [18]. Therefore, a given far field intensity distribution provides statistical information regarding the probability distribution of topological defects. We thus determined the probability of topological defects by fitting the experimental radial intensity profiles [23]. For the calculated profiles we superimposed an in-phase (q=0) solution and topological defect (q≠0) solutions, and varied the probability of these solutions to achieve the best fit to our data [35]. For 10 lasers, we obtained 2% topological defects, whereas for 20 lasers, the topological defects increased to 18% [43].

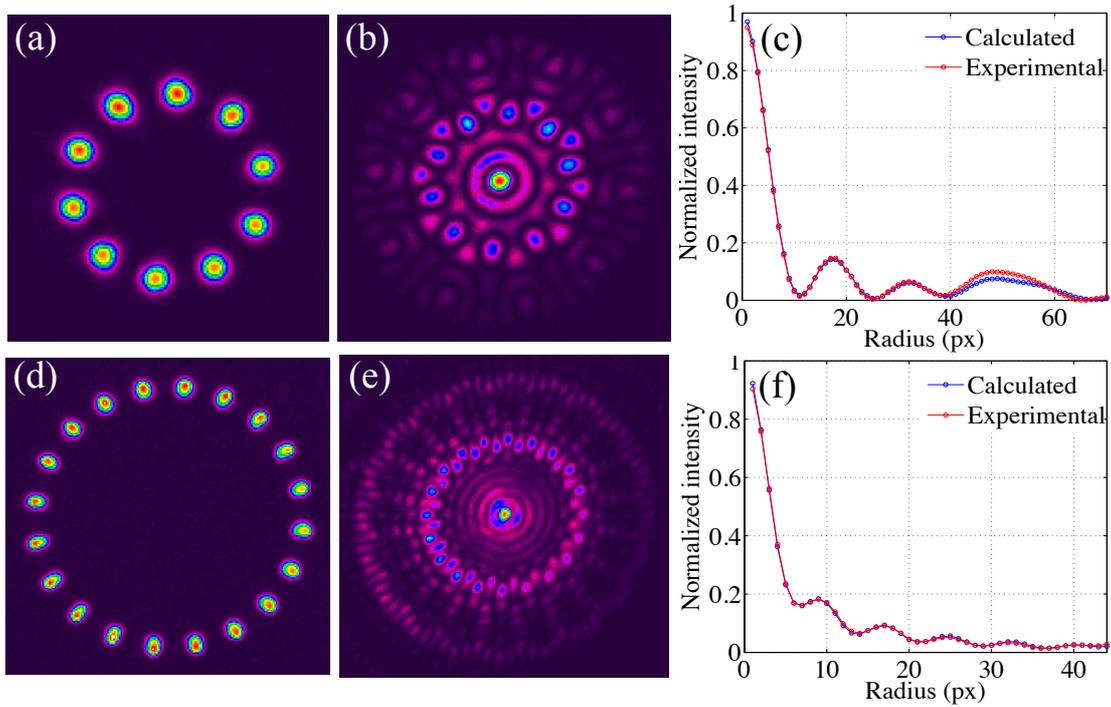

FIG 2. Topological defects in ring networks of 10 (top row) and 20 (bottom row) lasers. (a, d) Experimental near-field intensity distributions. (b, e) Experimental far-field intensity distributions after the SHG. (c, f) Experimental and calculated radial intensity profiles of far-field intensity distributions after SHG. The probability of topological defects is found to be 2% for 10 lasers, and 18% for 20 lasers.

To support our experimental results, we first integrate the Kuramoto equations (Eq. (4)) with initial phases randomly distributed between [0 to 2π], until a steady state solution was reached and determine its q. We repeated this for 5000 different random initial phases, and determined the probability of topological defects of globally stable (q=0) and locally stable (q≠0) solutions. The calculated probability of topological defects also increases with the system size [green squares in Fig. 3], but much higher than that obtained in the experiment.

The Kuramoto results can be explained by a simple physical point of view. First, for the initial conditions of random phases for N lasers located on a ring, the probability of topological defects can be derived from a "random walk" model [24] to yield a Gaussian distribution with variance ~ N [for large N [8]]. Second, the Kuramoto dynamics with static frequency disorder preserves the number of topological defects (as long as noises are small [22] so the same distribution is preserved at the steady state). Note, that in 2D geometry topological defects (vortices) are not fully protected, so the Kuramoto dynamics converges to a globally stable solution in un-frustrated geometries [18]. Alternatively, the Kuramoto model exhibits overdamped dissipative dynamics [22] over a complex energy landscape [19] and hence cannot get out of local minima. The distribution of topological defects hence represents the basin of attraction of the different locally stable solutions.

The large discrepancy between the Kuramoto results and our experimental results indicate that the coupled laser dynamics provides an additional mechanism to unwind topological defect and reach the globally stable solution more often than expected form the naive Kuramoto limit. To confirm and clarify such a mechanism, we resorted to the full laser rate equations [Eqs. (1)-(3)], which allow variation and fluctuation also of the lasers amplitudes. For $P_m/P_{th}$= 14.2 ($P_{th}$=1.2 is the threshold value), we obtain very good agreement between the simulated and the experimental results, as shown by the blue triangles in Fig. 3. This suggests that laser amplitudes dynamics indeed provide a mechanism to suppress topological defects.

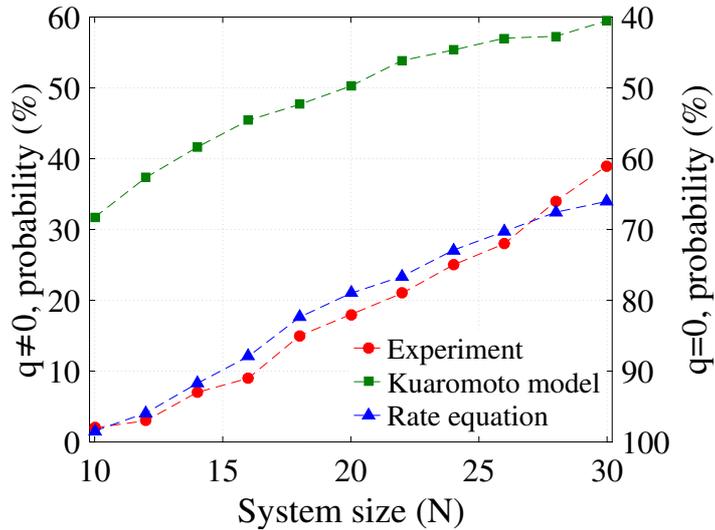

FIG. 3. Probability of locally stable topological defects (q ≠ 0) and globally stable in-phase solution (q = 0) as a function of system size. Red circles- experimental; blue triangles- numerical results with full laser rate equations that agrees well with the experimental data; green squares- numerical results with Kuramoto model that predict much larger defect probability than the experimentally measured one.

Since the amplitude dynamics is directly related to pump strength [44], we next studied the effect of pump strength $P_m$ on the probability of topological defects in steady state. For a ring network of 20 lasers, the experimental and simulated results, using the laser rate equations, are presented in Fig. 4. Figure 4(a) shows the radial profiles of the measured far-field intensity distributions. As evident, an increase in pump leads to greater smearing of the central dark and bright rings, indicating greater probability of topological defects. Fitting these profiles (as in Fig. 2), yields the quantitative probability of topological defects as a function of pump strength, as shown in Fig. 4(b). The simulated results of the laser rate equations agree well with the experimental results.

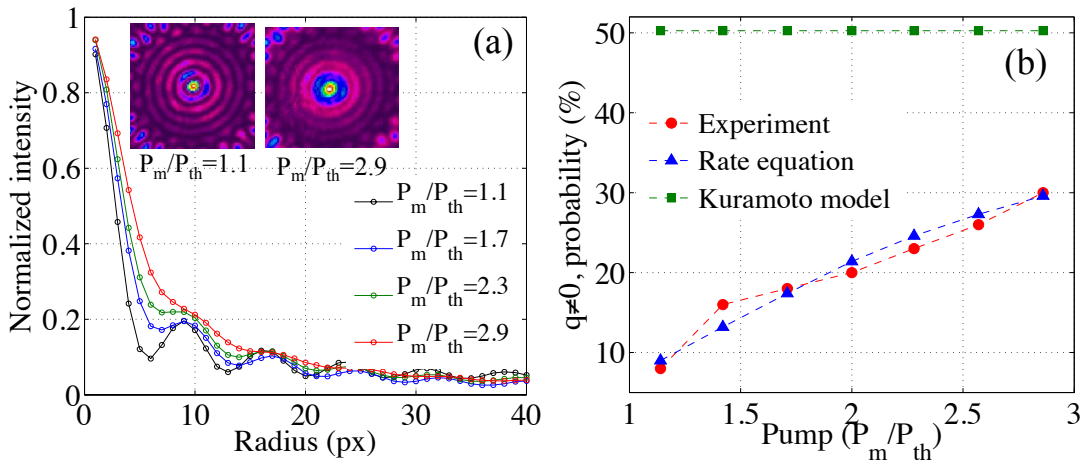

FIG. 4. Dissipative topological defects as a function of laser pump strengths in a ring network of 20 lasers. (a) The experimental radial intensity profiles for different normalized pump strengths ($P_m/P_{th}$). The increase in pump strength causes smearing of the dark and bright rings in the far-field intensity distributions, indicating an increased probability of topological defects. The insets show the far-field intensity distributions at $P_m/P_{th}$=1.1 and $P_m/P_{th}$=2.9. (b) Probability of topological defects as a function of normalized pump strength $P_m/P_{th}$. The blue triangles represents the simulated results of the laser rate equations that agree well with the experimental data (red circles) while the Kuramoto model (green squares) predicts much higher defect probability.

Furthermore, the probability of topological defects increases with the pump strength and approaches the Kuramoto results for $P_m/P_{th} \to \infty$, (when the Kuramoto assumption of uniform amplitudes is fulfilled [18, 28]). Alternatively, as $P_m$ approaches $P_{th}$, the effect of amplitudes dynamics becomes pronounced [44]. The probability of topological defects approaches zero and system dissipates to the globally stable solution. This agrees with recent results of an OPO-based Ising machine [19] where $P_m \sim P_{th}$ was shown to correspond to zero temperature, and $P_m > P_{th}$ was shown to correspond to a finite temperature [25].

The control over topological defects by the pump strength and analogy with temperature can be put in the context of KZ mechanism, where the density of defects scales with cooling rate at which the phase transition is crossed [2, 3]. In an analogy to Zurek's approach, we identify two competing time scales in our system, namely the amplitude synchronization time of lasers field amplitudes ($t_{amp}$) and the locking time of their phases ($t_{phase}$). To evaluate how these two competing time scales depend on various laser parameters, and to explore the underlying physical mechanism for the formation of topological defects, we simulated the dynamical behaviors of the field amplitudes and phases of the lasers using Eqs. (1)-(3).

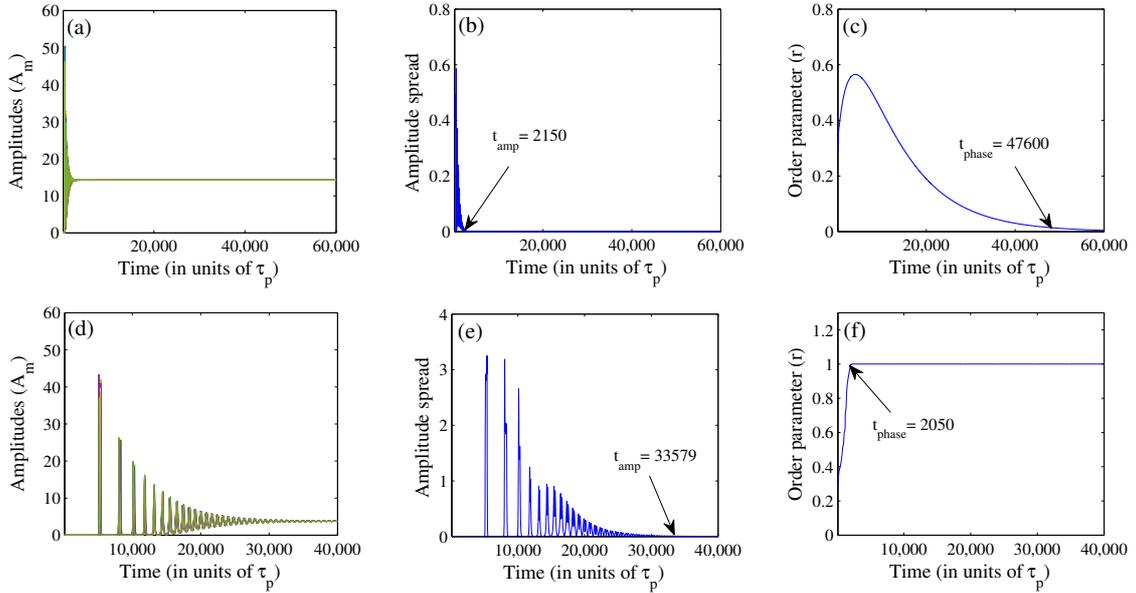

FIG 5. Simulated dynamical behaviors of the laser amplitudes and phases at two different pump strengths, for a ring network of N=20 lasers. The amplitudes, amplitude spread and phase order parameter are shown as a function of time for pump strength $P_m/P_{th}$=16.7 in (a), (b) and (c) and for $P_m/P_{th}$=1.25 in (d), (e) and (f) with the same initial conditions and coupling strength κ = 0.001. The marked arrows represent the amplitude synchronization time ($t_{amp}$) and phase locking time ($t_{phase}$). When $t_{amp} \ll t_{phase}$ the system get stuck in a locally stable state (top) with q ≠ 0, and when $t_{amp} > t_{phase}$ the system reaches the globally stable q = 0 state (bottom).

The results of these simulations at two different pump strengths, for a system of N=20 lasers, are shown in Fig. 5. Figures 5(a) and 5(d) show the amplitude dynamics of all 20 lasers at $P_m/P_{th}$=16.7 and at $P_m/P_{th}$=1.25. As evident, for high $P_m/P_{th}$ the amplitudes $A_m$ initially fluctuate but reach a steady-state value very fast, whereas for low $P_m/P_{th}$ the amplitudes reach steady-state much slower. The resulting (normalized) amplitude spread among the lasers $\sqrt{\langle (A_m - \langle A_m \rangle)^2 \rangle}/\langle A_m \rangle$, shown in Figs 5(b) and 5(e) follows the same trend. In particular, the time $t_{amp}$ where the envelope of amplitude spread reaches 0.01 is ~15 times longer for low $P_m/P_{th}$. Figures 5(c) and 5(f) show the phase locking dynamics, characterized by the order parameter $r(t) = \left| \frac{1}{N} \sum_{m=1}^{N} e^{i\varphi_m(t)} \right|$ [22]. For high $P_m/P_{th}$ phase locking is much slower than for low $P_m/P_{th}$ and $t_{phase}$ (the time where $r(t)$ reaches within 0.01 of its steady state value) is ~23 times longer. In summary, at high $P_m/P_{th}$, $t_{amp} \ll t_{phase}$, whereas at low $P_m/P_{th}$, $t_{amp} \gg t_{phase}$.

The amplitude fluctuations can be regarded as an effective temperature [28] that is then coupled to the phase dynamics [see Eq. (2)] and the phase locking time t_phase corresponds to the information exchange rate. Hence, for $t_{amp} \ll t_{phase}$ the "external" cooling rate is much faster than the "internal" information exchange rate, thereby promoting the generation of topological defects according to the KZ picture. For $t_{amp} \gg t_{phase}$ the fluctuating amplitudes can "kick" the system out of locally stable solutions, allowing it to find the globally stable solution.

To quantify and generalize the relation between topological defects and the competing time scales of the system, we analyzed the probability of topological defects in the network also for different coupling strengths. For each coupling strength, we varied the pump and calculated the two time scales $\langle t_{amp} \rangle$ and $\langle t_{phase} \rangle$, averaged over 1000 random initial conditions. The results are shown in Fig. 6. As evident, the probability of topological defects grows as $\langle t_{amp} \rangle$ *decreases* (Fig. 6(a)) and as $\langle t_{phase} \rangle$ *increases* (Fig. 6(b). When the probability of topological defects is plotted as a function of $\langle t_{phase}^{0.6} \rangle / \langle t_{amp} \rangle$, all trajectories collapse to a single curve, as shown in Fig. 6(c). Such data collapse indicates universality in the system, although we cannot explain the value of the exponent theoretically. The increase of probability of topological defects with the increase in the ratio between $\langle t_{phase} \rangle$ and $\langle t_{amp} \rangle$ agrees with the KZ picture.

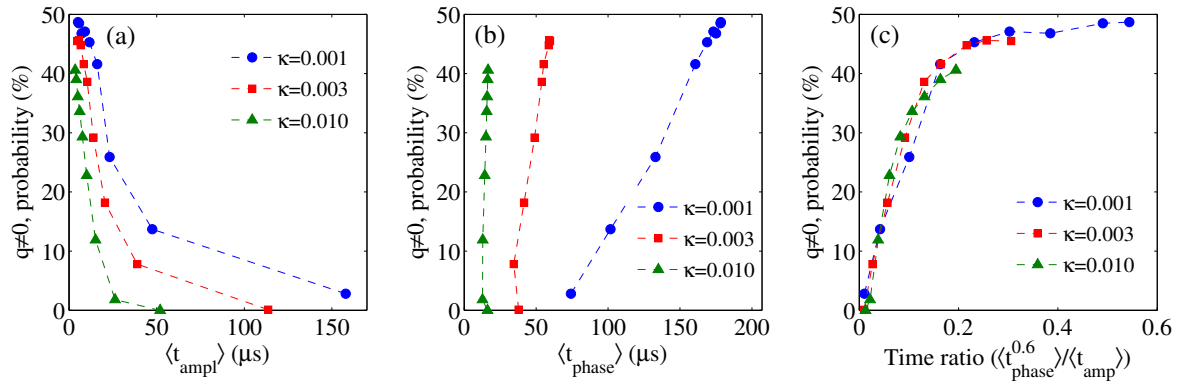

FIG. 6. Simulated probability of topological defects, using the laser rate equations, for a ring network of 20 lasers (a) as a function of synchronization time of laser amplitude fluctuations at three different coupling strengths, (b) as a function of phase locking time at different coupling strengths, and (c) as a function of the ratio $\langle t_{phase}^{0.6} \rangle / \langle t_{amp} \rangle$. In all figures, for each coupling strength, every point on the curve corresponds to different pump strength. All trajectories collapse to a single curve, indicating an universality in the system.

Finally, we present an even closer analogy of the KZ mechanism using the Kuramoto model [Eq. (4)] and sweeping the coupling strength $\kappa(t)$ between the lasers across the phase transition at different rates. Our numerical results, presented in Fig. 7, show that the probability of topological defects grows with the sweeping rate of the coupling strength across the phase transition from unlocked to the phase locked state, again in agreement with the KZ picture.

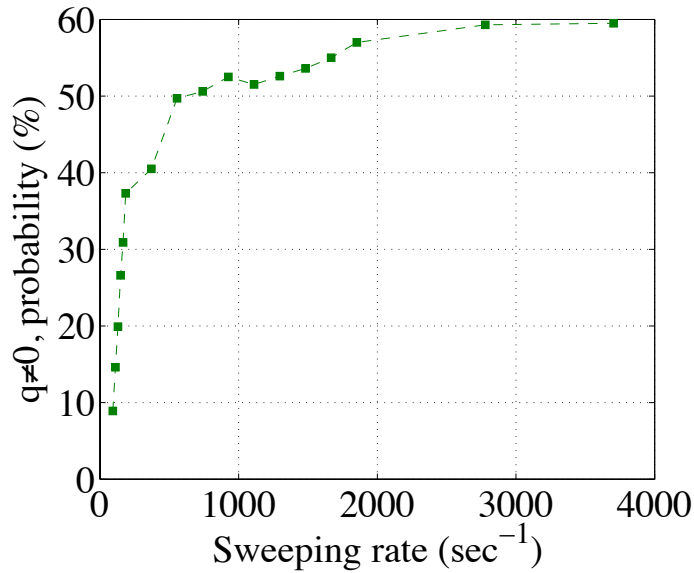

FIG. 7. The probability of topological defects, as a function of sweeping rate of coupling strength $\kappa(t)$ (from 0.0001 to 0.35) for 30 lasers network.

**CONCLUDING REMARKS**

We have investigated dissipative topological defects in a 1-D ring laser networks, and showed how they are linked to the KZ mechanism. The probability of topological defects increases with the system size and the pump strength. The formation of topological defects depends on two competing time scales, namely phase locking time and synchronization time of amplitude fluctuations. When the phase locking (information exchange) time is much shorter than the synchronization (cooling) time, the probability for topological defects formation approaches zero in agreement with the KZ picture. We observed universality in the system where the probability of topological defects depends only on a single parameter related to the ratio between the two competing time scales. As opposed to the KZ mechanism where the system is cooled by an external heat bath, in our ring laser network the amplitude fluctuations act as an "internal" heat bath that is coupled to the phases.

VP thanks for the financial support from Planning and Budgeting Committee (PBC) Postdoctoral Fellowship.

Supplementary material for

# Dissipative Topological Defects in Coupled Laser Networks

Vishwa Pal, Chene Tradonsky, Ronen Chriki, Asher A. Friesem and Nir Davidson.

**Details about least square fitting to extract the probability of TC**

In the experiment, the collective ensemble measurements of the lasers outputs were comprised of both globally stable in-phase solution and locally-stable topological defects. To extract the percentage of different solutions, we fit the experimental results, using least square method, with the calculated results. For the calculations we assumed that the laser output intensities are equal and have Gaussian profiles, and the output distributions are out-of-phase or in-phase when the lasers are phase locked. After arranging the Gaussian fields on ring geometry and performing a Fourier transform, we obtained the calculated far-field intensity distributions.

The phases of the Gaussian fields were chosen in in-phase configuration (0, 0, 0, 0...) for globally stable solutions, and in topological defect configuration ($\varphi_m = e^{iqm2\pi/N}$) for locally stable solutions, where q represents the winding number of topological defects and *m* denotes the index of Gaussian field on the ring array. We verified numerically, using Eqs. (1)-(3), that these phase distributions are indeed stable stationary solutions with nearest neighbour instantaneous coupling and parameters suitable for our experimental system. The calculated results of far-field intensity distributions and their corresponding radial profiles for two different size ring networks are shown in Figs. S1 and S2.

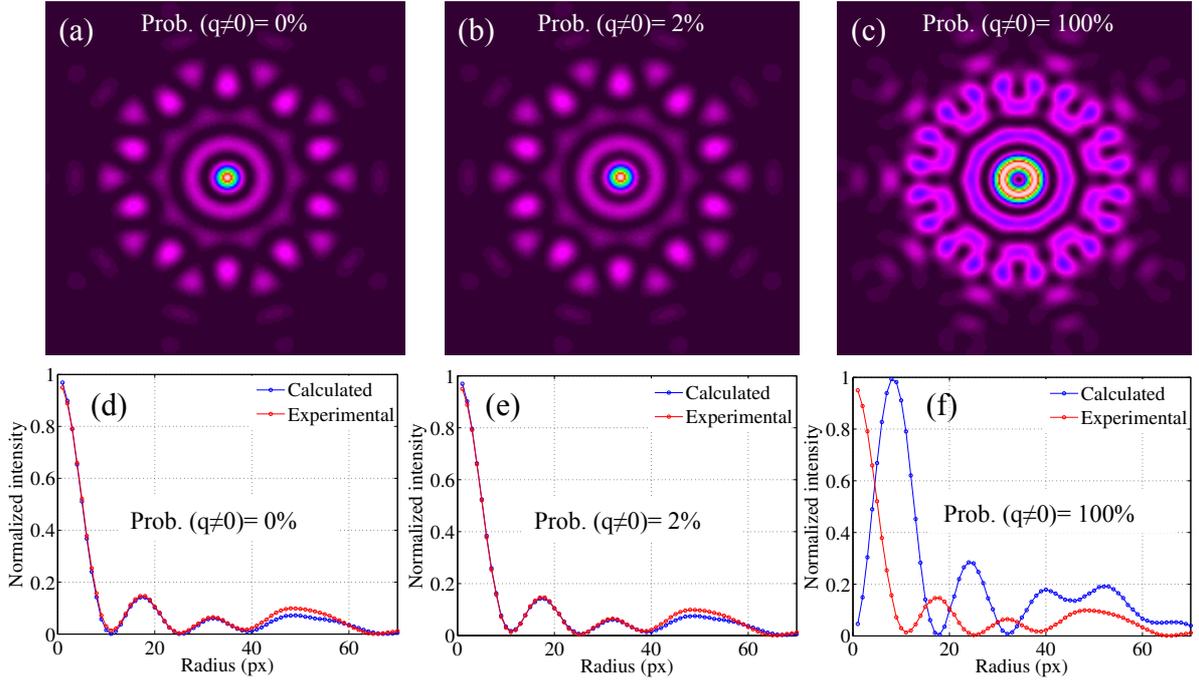

FIG. S1: Calculated far-field intensity distributions (a-c) and their radial profiles (d-f) for 10 lasers ring network, with different probability of topological defects: (a, d) 0%, (b,e) 2%, and (c,f) 100%. Note that here the order of |q| was chosen as 1.

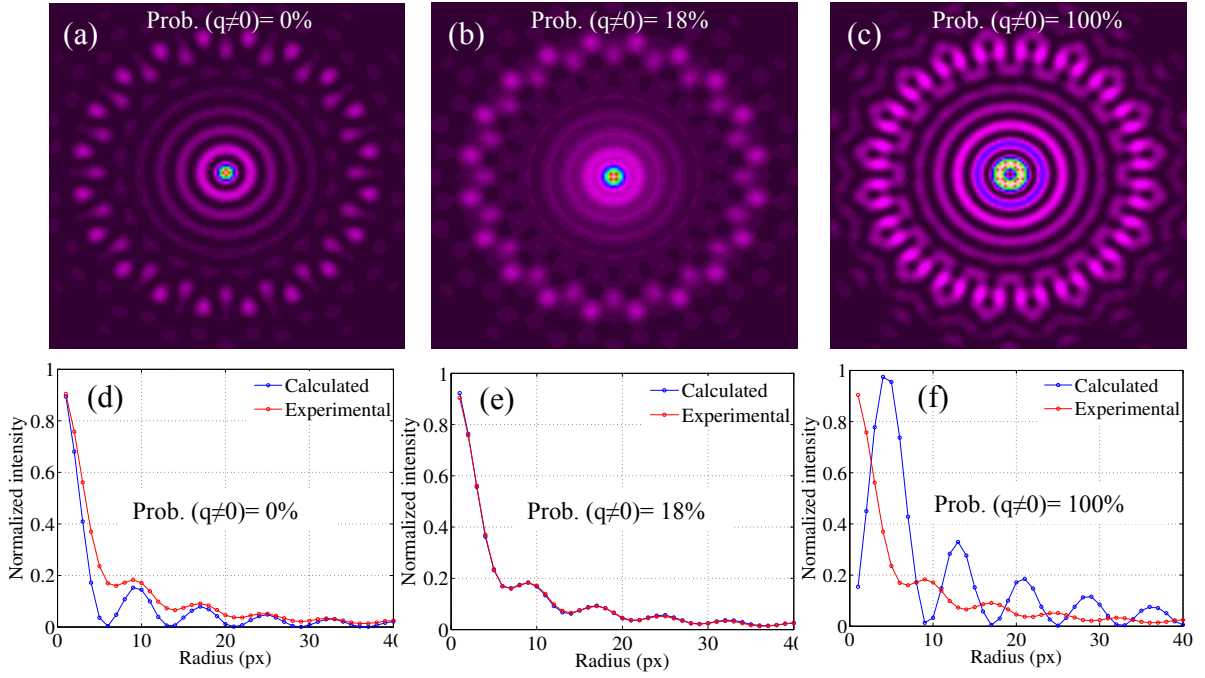

FIG. S2: Calculated far-field intensity distributions (a-c) and their radial profiles (d-f) for 20 lasers ring network, with different probability of topological defects: (a, d) 0%, (b,e) 18%, and (c,f) 100%. Note that here the order of |q| was chosen as 1 and 2.

**Phase locking analysis**

To converge our ring laser network to a globally or locally stable solution, it is necessary that all lasers be phase locked and obey periodic boundary conditions. We verified that indeed all lasers were phase locked in our experiments by interfering one laser with all other lasers, as schematically shown in Fig. S3.

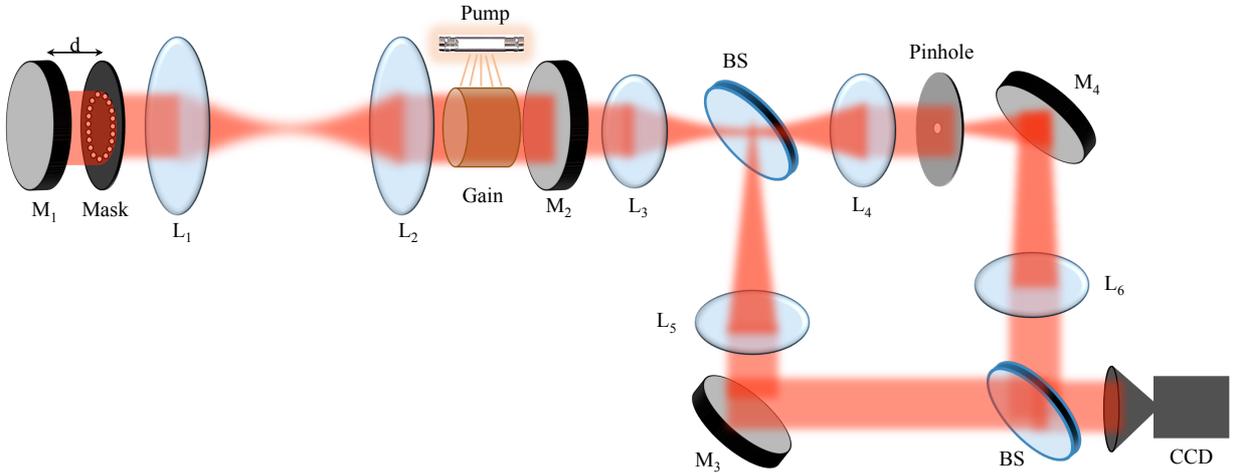

Fig. S3: Degenerate cavity for obtaining phase locked lasers in a ring geometry, and Mach-Zehnder interferometer arrangement to analyze the phase locking between the lasers. $M_1$ and $M_2$: high reflectivity and partially reflecting cavity mirrors, respectively; $M_3$ and $M_4$: high reflectivity mirrors; $L_1$, $L_2$, $L_3$, $L_4$, $L_5$, $L_6$: lenses; BS: beam splitter; CCD: detection camera.

In the interferometer, the phase locked array of lasers splits into two arms. In one arm the output from the degenerate laser is imaged onto the CCD camera. In the second arm, the light from a single laser is selected with a pinhole of 50 μm diameter, and then expanded in order to overlap with the light from all the lasers at CCD camera. Accordingly, the light from a single laser interferes with itself and with the light from all the other lasers. A phase tilt between the light in the two arms is introduced by an angular orientation of second beam splitter, so the fringes are formed. The CCD detects the resulting interference pattern.

Representative interference patterns for different sizes of ring laser networks are shown in Fig. S4. As evident, fringes appear at each spots, which confirms that all of the lasers in the network are phase locked. However, the fringe contrast degrades as further away from the reference laser, and this is more prominent in the large networks.

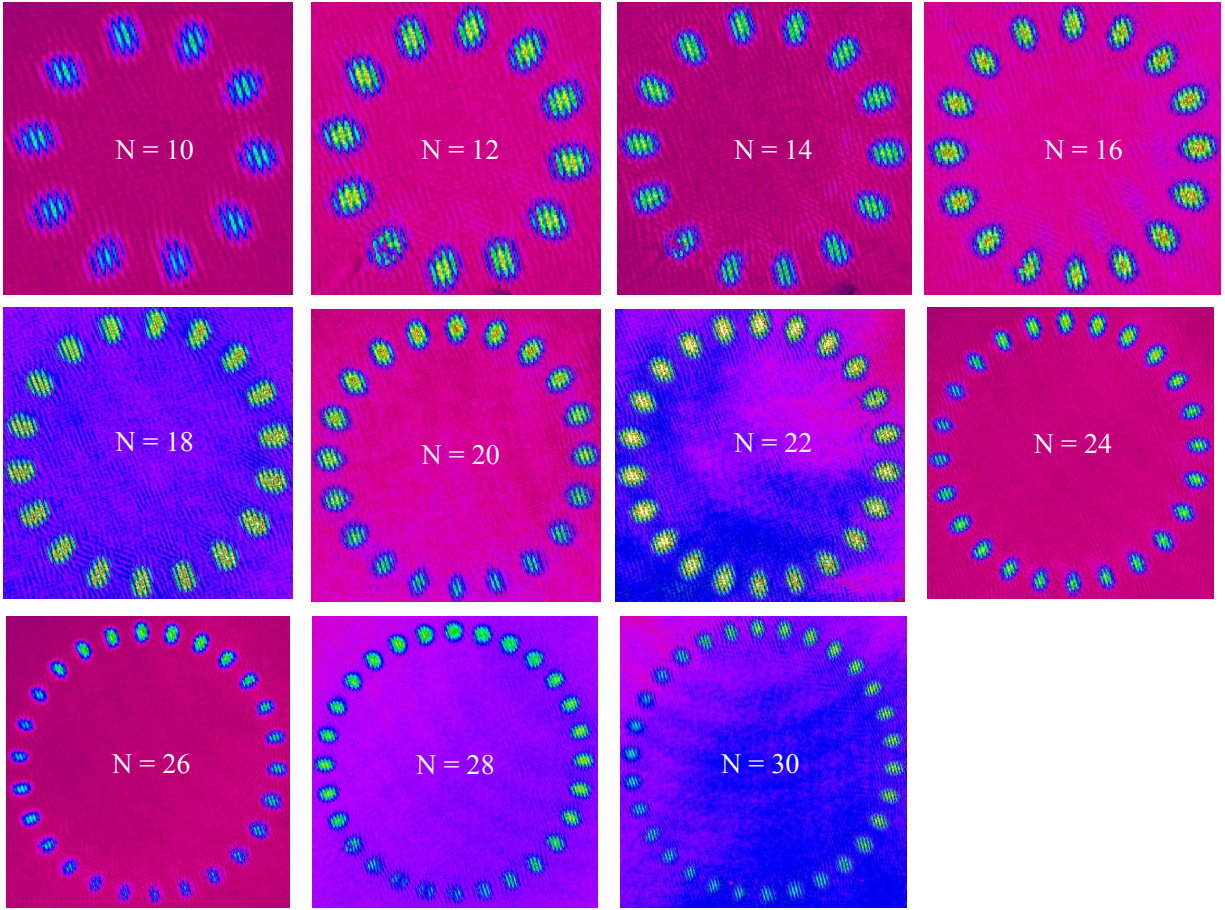

FIG. S4: Interference patterns for different size ring laser networks.

**Eigenvalue analysis for different sizes of 1-D ring laser networks**

We calculated and compared the pump threshold values of globally and locally stable solutions using Eigenvalue analysis, for ring laser network of different sizes. The results are shown in Fig. S5. The solutions with low pump threshold values are more probable than those with the higher threshold values. For a given network size, the pump threshold values are lowest for globally stable solutions (q=0), and increase with the winding number of locally stable solutions (q≠0). Moreover, the difference between the pump thresholds of globally and locally stable solutions depends on the network size, as shown in Fig. S5. For small size networks, the difference is large, which implies that system can easily dissipate to the globally stable solution. Whereas, for large size networks the difference is small, so the probability of dissipation to the globally stable solution goes significantly down, and system gets stuck in the locally stable solutions.

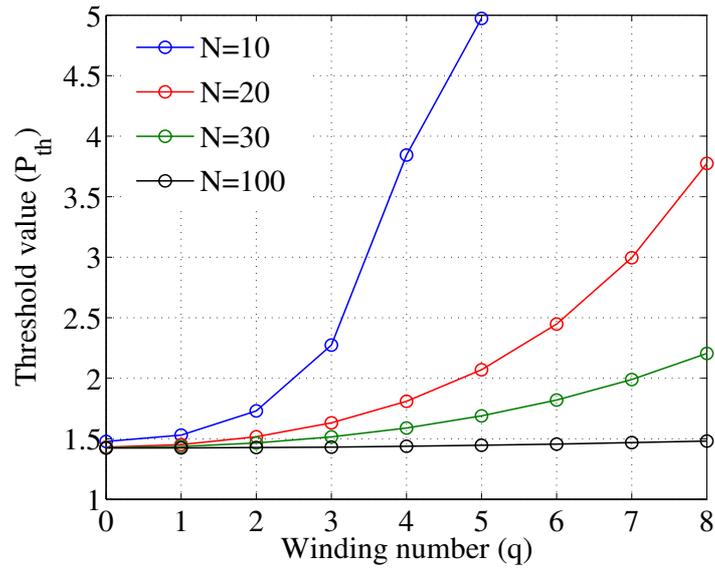

Fig. S5: Comparison of pump threshold values of globally and locally stable solutions for laser ring networks of different sizes N=10, 20, 30 and 100.